\documentclass[journal]{IEEEtran}
\IEEEoverridecommandlockouts

\usepackage[hidelinks]{hyperref}
\usepackage[cmex10]{amsmath}
\usepackage{amssymb,amsfonts}
\usepackage{dblfloatfix}

\usepackage[ruled,vlined]{algorithm2e}
\usepackage{graphicx}

\usepackage{booktabs}
\usepackage{siunitx}
\usepackage[numbers,compress]{natbib}
\usepackage{texnames}
\usepackage{bm,bbm}
\usepackage{orcidlink}
\usepackage{subcaption} 
\usepackage{multirow}
\usepackage{placeins} 

\begin{document}

\title{Onboard Hyperspectral Super-Resolution with Deep Pushbroom Neural Network}

\author{Davide Piccinini\orcidlink{0009-0000-9675-8008},~Diego~Valsesia,~and~Enrico~Magli\\
\small Department of Electronics and Telecommunications, Politecnico di Torino, Italy\\
\small \texttt{\{name.surname\}@polito.it}%
}


\maketitle
\begin{abstract}
Hyperspectral imagers on satellites obtain the fine spectral signatures essential for distinguishing one material from another at the expense of limited spatial resolution. Enhancing the latter is thus a desirable preprocessing step in order to further improve the detection capabilities offered by hyperspectral images on downstream tasks. At the same time, there is a growing interest in deploying inference methods directly onboard of satellites, which calls for lightweight image super-resolution methods that can be run on the payload in real time. In this paper, we present a novel neural network design, called Deep Pushbroom Super-Resolution (DPSR), which matches the pushbroom acquisition of hyperspectral sensors by processing an image line-by-line in the along-track direction with a causal memory mechanism to exploit previously acquired lines. This design greatly limits memory requirements and computational complexity, achieving onboard real-time performance, i.e., the ability to super-resolve a line in the time it takes to acquire the next one, on low-power hardware. Experiments show that the quality of the super-resolved images is competitive or even outperforms state-of-the-art methods that are significantly more complex. Code and pretrained models are available at: \url{https://github.com/DavidePiccinini98/dpsr}.
\end{abstract}


\section{Introduction}
\IEEEPARstart{H}{yperspectral} imagery has emerged as a vital tool in the field of remote sensing because it samples a scene across a vast number of narrow spectral bands. This fine spectral resolution allows analysts to precisely identify and characterize the unique spectral signatures of different materials. Consequently, hyperspectral images are proving invaluable in a broad range of applications, including environmental conservation, agricultural management, pollution control, mineral exploration. By examining the subtle reflectance properties of surfaces, e.g. computing vegetation health indicators, soil composition, or building materials, hyperspectral imaging supports enhanced data-driven decision-making and accurate assessment. 

Despite these distinct advantages, the benefits of hyperspectral imaging come with certain trade-offs. One of the most significant drawbacks lies in its typically coarser spatial resolution when compared to multispectral sensors. While these easily provide spatial resolutions in the order of few meters per pixel, hyperspectral payloads are limited to tens of meters per pixel or worse. For instance, the recent PRISMA \cite{cogliati2021prisma}, EnMAP \cite{9484000} and HySIS \cite{eoportal_hysis} missions all provide a resolution of 30 m/pixel.
For this reason, spatial super-resolution of hyperspectral images is a topic of great interest. Recent literature \cite{chen2023msdformer, xu2024hyperspectral, wang2024ssaformer, hu2024exploring}, particularly using deep neural networks, has shown that powerful data priors can be used to accurately enhance spatial resolution, even from a single image.

At the same time, there is a growing interest in remote sensing community to shift image analysis procedures from ground-based operations to onboard satellite platforms \cite{giuffrida2021varphi, ruuvzivcka2022ravaen, ziaja2021benchmarking, inzerillo2024efficient}.
By processing data directly in space, as they are acquired, satellites can rapidly identify and respond to critical occurrences such as natural disasters or sudden environmental changes. This approach significantly reduces the latency associated with transferring large volumes of raw data to Earth, followed by subsequent processing at ground stations. Ultimately, real-time or near-real-time analytics could provide timely information to decision-makers and emergency response units, potentially saving lives and mitigating damage. However, onboard processing faces significant constraints in terms of available computational resources, which calls for the development of efficient and low-complexity models. Hyperspectral imagers represent a good case study since they generate very large volumes of spatial-spectral data which may be difficult to process efficiently, particularly in terms of memory requirements.

Since super-resolution of hyperspectral images is highly desirable for more accurate image processing in downstream applications, it is natural to wonder whether efficient models could be developed so that a satellite payload could perform super-resolution in real-time as it acquires images. On the other hand, the current literature on hyperspectral super-resolution \cite{chen2023msdformer, chen2024cross} predominantly emphasizes maximizing output quality, at the expense of complexity, often relying on sophisticated and computationally expensive deep learning architectures. While these models achieve impressive results, they are ill-suited for potential onboard usage, where constraints related to power, memory, and processing speed are stringent. Concretely, current high-performance models are orders of magnitude heavier than what a small, low-power accelerator can sustain. For example, on the HySpecNet-11k dataset, state-of-the-art methods MSDformer \cite{chen2023msdformer} and CST \cite{chen2024cross} require $714$K and $245$K FLOPs per pixel (FLOPs/px), respectively, for $4\times$ super-resolution, and $528$K and $121$K FLOPs/px for $2\times$ super-resolution, which are excessive for existing low-power accelerators. Under a realistic input of $1000  \times 1000  \times 66$, such as the one of the PRISMA VNIR instrument, they also exceed 24 GB of required memory, again precluding onboard use.   
Balancing the need for high-fidelity enhancements with the practical limitations of satellite hardware remains a challenging area of study.

In this paper, we depart from the literature and design a highly efficient model for hyperspectral image super-resolution for onboard usage, called Deep Pushbroom Super-Resolution (DPSR), that can run in real time on low-power hardware and with limited memory requirements, while providing performance comparable to that of state-of-the-art methods exploiting one order of magnitude more floating point operations (FLOPs) per pixel. On HySpecNet-11k with a super-resolution factor of $4\times$, DPSR only requires $31$K FLOPs/px, while with a factor $2\times$ $20$K FLOPs/px. Compared to other lightweight models such as EUNet \cite{liu2023efficient} and SNLSR  \cite{hu2024exploring}, the core of our contribution is to have a neural network that sustains the acquisition dynamic of a pushbroom sensor by processing one line at a time as it is acquired. A memory mechanism based on Selective State Space Models (SSMs, e.g., Mamba \cite{gu2023mamba}) processes the image as a sequence of lines (as opposed to a 2D tile), and ensures that information from previously acquired lines is exploited to super-resolve the current line. This design minimizes the memory requirements since only the feature maps for the current line (and a small memory state) need to fit in the accelerator memory, instead of buffers of hundreds of lines and respective features needed by state-of-the-art methods. We show that DPSR can process an entire PRISMA VNIR frame of size $1000 \times 1000  \times 66$ with less than $1$ GB of memory, at least an order of magnitude less than other existing methods. Notice that this line-based paradigm naturally matches the pushbroom imaging method commonly used in hyperspectral sensors on satellites, which acquires one line with all its across-track pixels and spectral channels at a given time, and uses the movement of the satellite in the along-track direction to capture successive lines. Consequently, our super-resolution module could be directly pipelined to the sensor output, enabling real-time image enhancement as a line is super-resolved in the acquisition time of the next one without the need for extra line buffering. As an example, the line acquisition time for the PRISMA satellite is 4.34 ms \cite{prismaATBD} and our experimental results show that a largely unoptimized implementation of DPSR on a 15 W system-on-chip super-resolves a line\footnote{$2\times$ scaling factor, thus producing 2 output lines with twice as many across-track pixels; the line size is the one of the VNIR PRISMA instrument, i.e. $1\times 1000 \times 66$.} in 4.25 ms, demonstrating, for the first time, real-time performance. On the other hand, we show that existing state-of-the-art models in the literature, including those more efficiency-oriented such as EUNet \cite{liu2023efficient} and SNLSR  \cite{hu2024exploring}, all run far slower than the real-time threshold of $4.34$ ms (see Figures \ref{fig:speed_x2} and \ref{fig:speed_x4}).

\begin{figure*}[t]
    \centering
    \includegraphics[width=0.9\textwidth]{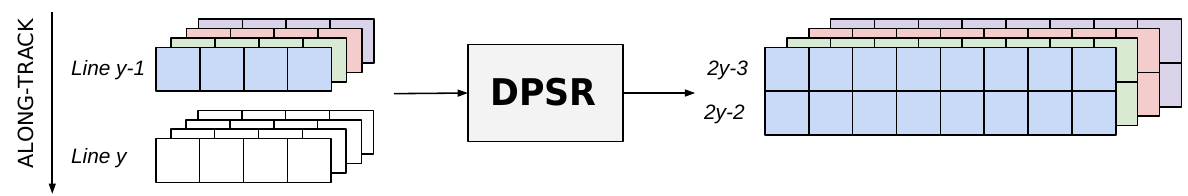}
    \caption{DPSR spatially super-resolves a hyperspectral image line-by-line, by using line $y-1$ and line $y$, as well as a compact memory of past lines to super-resolve line $y-1$ in the along- and across-track directions.}
    \label{fig:method}
\end{figure*}

A preliminary version of this work appeared in \cite{piccinini2025LineSR}. The present paper substantially extends and improves upon that earlier version in both methodology and evaluation. Specifically, the model architecture has been redesigned in several ways. First, we now treat the spectral dimension as a feature axis rather than as part of a 3D spatial cube, leading to a more efficient and expressive representation. Moreover, a new design of the basic neural block (SFE Block in this paper) uses attention operations rather than simple convolutions, and a residual connection with bilinear interpolation is introduced. Following these modifications, substantial improvements have been obtained, with DPSR delivering a $43.17$ dB MPSNR on HySpecNet-11k ($4\times$ SR factor), compared to the $41.82$ dB achieved by LineSR. This paper also significantly expands the discussion and the experimental validation, providing results under two super-resolution factors ($2\times$ and $4\times$) across four different datasets, comparing against seven baselines and showing extensive ablation studies that dissect the contributions of individual architectural components and design choices. Finally, experiments on low-power hardware are also added to support the low-complexity claims and suitability for onboard deployment. 

We can summarize our contributions as follows:
\begin{itemize}
\item We introduce a lightweight deep learning architecture tailored for hyperspectral single-image super-resolution onboard satellites.
\item Drawing inspiration from the operational principles of pushbroom sensors, our framework performs super-resolution in a causal, line-by-line fashion. To accomplish this, we leverage deep SSMs, which maintain an effective memory of previously processed lines. This design choice allows the network to access relevant historical features without storing or reprocessing the entire image, leading to substantial efficiency gains.
\item Experimental results on multiple datasets show image quality comparable to state-of-the-art models at a fraction of complexity, and significantly lower runtime and memory requirements.
\end{itemize}

\section{Background}

\subsection{Super-resolution of hyperspectral images}
Hyperspectral image super-resolution stands out as a pivotal challenge in the remote sensing community. The natural trade-off between spectral and spatial resolution often leaves hyperspectral data with relatively coarse pixel sizes. Enhancing the spatial detail of hyperspectral data opens the way to more accurate image analysis in downstream tasks such as land cover mapping, vegetation health monitoring, mineral deposits detection, and many more. In this paper, we focus on single-image super-resolution where only one image at the lower resolution (LR) is available. Multi-image methods \cite{valsesia2021permutation} exploiting multi-temporal data are known to better regularize the problem and further enhance quality. However, our scenario is that of super-resolving the image immediately as it is acquired by the instrument, which is inherently a single-image task. 

Historically, researchers have explored a wide array of approaches to tackle this task. Early solutions relied on classical signal processing and machine learning techniques, such as low-rank and sparse coding \cite{huang2014super}, spectral-spatial sub-pixel mapping \cite{xu2016hyperspectral}, and low-rank tensor decomposition integrated with total variation regularization \cite{he2016super}. While these methods demonstrated good results, they generally depended on handcrafted priors, which could be challenging to adapt across different sensors or environmental conditions.

The advent of larger hyperspectral datasets and the parallel rise of deep learning have significantly reshaped the field. In particular, convolutional neural networks (CNNs) have become a popular tool \cite{ mei2017hyperspectral, arun2020cnn} for super-resolution because of their powerful feature extraction capabilities and their inherent capacity to learn both spectral and spatial correlations directly from training data. These CNN-based models have substantially outperformed traditional methods, spurring further research to refine network architectures and training strategies. 
In \cite{li2018single} (GDRRN), the authors proposed a deep architecture based on the recursive use of residual connections and blocks comprised of grouped convolutions and nonlinearities to obtain super-resolved images.    

Subsequently, attention-based mechanisms introduced new ways to capture dependencies across spatial and spectral dimensions. Recent hybrid models \cite{jiang2020learning, liu2022interactformer, liu2023efficient, zhang2023essaformer, chen2023msdformer, wang2024ssaformer} have sought to combine the local feature extraction strengths of CNNs with the global context modeling capabilities of Transformers. By leveraging attention to establish relationships between distant spatial locations or widely separated spectral bands, these models have pushed performance even further. 
In \cite{jiang2020learning} (SSPSR), the authors proposed a deep architecture based on progressive upsampling, with parameters shared between branches of the network, and use of channel attention to fully capture spectral information. In \cite{liu2023efficient} (EUNet), a multi-stage network that uses both a degradation model constraint and a deep learning-based super-resolution model is proposed, aiming at iteratively refining the SR image obtained; their strategy is based on splitting the MAP
optimization problem into two subproblems, i.e., data consistency and prior terms, and then jointly optimize them. At the same time, \cite{chen2023msdformer} (MSDformer), proposed a Transformer-based architecture comprising two modules to capture both spectral and spatial information: a Multiscale Spectral Attention Module that makes use of a novel spectral grouping strategy and a combination of convolutions and channel attention, and a Deformable Convolution-Based Transformer Module that uses a modified self-attention block that has feature maps obtained with convolutions.

More recently, in \cite{hu2024exploring} (SNLSR), the authors introduced a very efficient deep learning SR method that first unmixes the LR image into its abundance and endmember parts, then it super-resolves the LR abundance to get its representation in the high-resolution space and finally, the super-resolved abundance is linearly mixed with the same endmembers. Their work is based on the hypothesis that HSI can be efficiently decomposed into a linear spectral combination of its abundance
fractions and endmembers \cite{wang2022dilated}. In \cite{Dong18072025} an effective combination of Long Short-Term Memory networks (LSTMs) and multi head attention delivering competitive results was presented. In \cite{chen2024cross}, the authors proposed a new Cross-scope Spatial-spectral Transformer (CST) model based on two modules they designed: a Cross-scope Spectral Self-attention Module and a Cross-scope Spatial one, which are built on top of window-based attention architecture. With these modules the authors wanted to extract both short and long range information in spatial and spectral directions obtaining richer representations of the input. Recently, diffusion-based generative SR methods have been extended to hyperspectral imagery \cite{Cheng22012025, 10644098}, albeit with considerably high computational cost. Lastly, new remote sensing foundation models have also been tested on the hyperspectral SR \cite{10949864} task, suggesting that foundation-scale pretraining can benefit HSI-SR.  

\subsection{Sequence Modeling} \label{sec:sequencemodeling}
Since the present work builds upon recent developments in sequence modeling, it is useful to review the main approaches adopted in deep learning for processing sequential data and to highlight their respective strengths and limitations.
A first classical approach to sequence modeling leverages causal convolutional layers, restricting each kernel to information from preceding time steps in order for the autoregressive flow to remain intact \cite{van2016pixel}.
A more effective method for modeling long sequences is based on Recurrent Neural Networks (RNNs): 
these models operate by recursively updating a hidden state as new inputs arrive, enabling them to capture temporal dependencies over time. RNN-based architectures have shown their effectiveness across a variety of tasks, including language modeling and time-series forecasting. However, a key limitation arises when dealing with long sequences, where the models may suffer from vanishing or exploding gradients. 
Making use of gating mechanisms, such as gates in LSTM networks \cite{hochreiter1997long} and Gated Recurrent Units \cite{cho2014properties, chung2014empirical}, alleviates but does not completely remove this bottleneck, and the strictly step-by-step computation makes it hard to harness the full parallelism of modern hardware when training on massive data sets.

Transformers \cite{vaswani2017attention} have recently replaced recurrent networks as the go-to architecture for sequence modeling. Instead of a step-by-step recursion mechanism, they make use of attention, which lets the model weigh every position in a sequence against every other, adapting its computation to the specific input and capturing dependencies no matter how far apart. Because each layer can be trained in parallel, Transformers scale far better than RNNs and are able to excel in tasks as varied as large-scale text \cite{team2023gemini, touvron2023llama}, image \cite{hatamizadeh2024diffit} and video generation \cite{brooks2024video}, and even protein-structure prediction \cite{jumper2021highly}.

\begin{figure*}[t]
    \centering
    \includegraphics[width=0.7\textwidth]{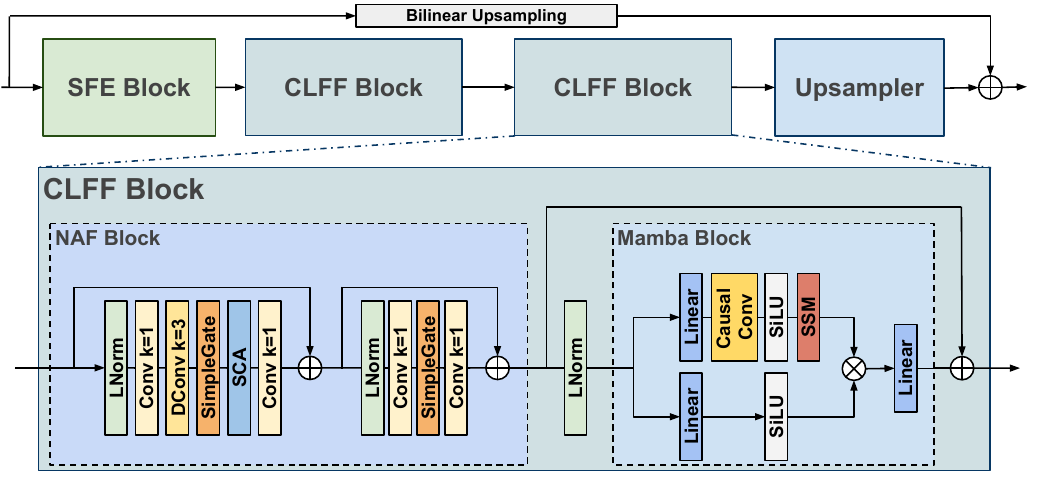}
    \caption{DPSR processes a line with an initial SFE Block followed by two CLLF Blocks composed of a NAFBlock to exact across-track and spectral features and a Mamba Block to provide context from previous lines. The architecture works as a residual correction to bilinear upsampling.}
    \label{fig:architecture}
\end{figure*}

\begin{figure*}[t]
    \centering
    \begin{subfigure}{0.60\textwidth}
        \centering
        \includegraphics[width=0.9\linewidth]{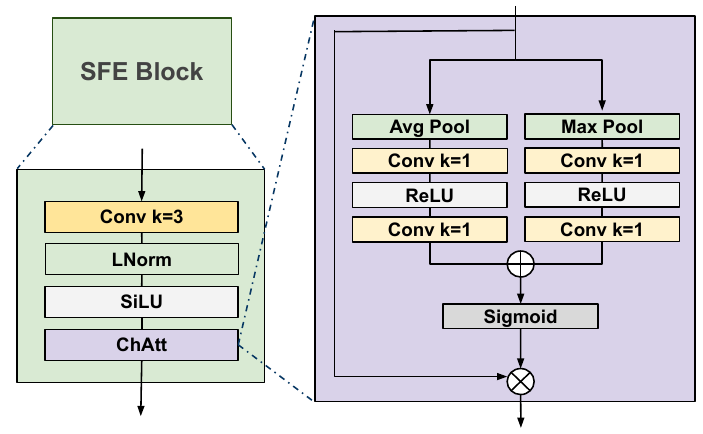}
        \label{fig:SFE}
    \end{subfigure}
    \hfill
    \begin{subfigure}{0.38\textwidth}
        \centering
        \includegraphics[width=0.9\linewidth]{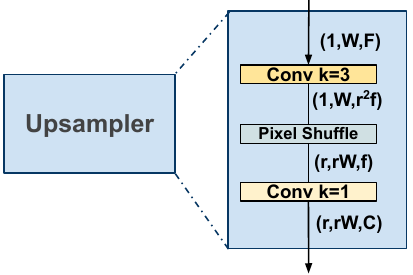}
        \label{fig:Upsampler}
    \end{subfigure}
    \caption{Left: architecture of SFE Block. Its main purpose is to filter the initial input in order to remove noise ad extract shallow features that will be later refined by the subsequent CLFF Block. Right: architecture of Upsampler module. It leverages the Pixel Shuffle \cite{shi2016real} and two convolutions to efficiently manage complexity.}
    \label{fig:SFE+Up}
\end{figure*}

The main drawback of Transformers is that self-attention grows quadratically in cost and memory with sequence length, a serious problem for very long inputs or latency-sensitive, memory-constrained applications. Researchers have proposed remedies to tackle the problem, such as sparse \cite{child2019generating} and linear attention \cite{katharopoulos2020transformers} variants, but fully resolving these scalability limits remains an open challenge.

More recently, a new class of models based on State Space Models (SSMs) has emerged as a compelling alternative for sequence modeling. These models are designed to process long sequences efficiently by modeling the temporal dynamics through a latent state vector that evolves over time. Formally, an SSM is defined by a set of equations:

\begin{align}
    \mathbf{h}'(t)&=\mathbf{A}\mathbf{h}(t)+\mathbf{B}\mathbf{x}(t),\\
    \mathbf{y}(t)&=\mathbf{C}\mathbf{h}(t)+\mathbf{D}\mathbf{x}(t),
\end{align}

where $\mathbf{x}(t) \in \mathbb{R}^M$ denotes the input at time $t$, $\mathbf{h}(t) \in \mathbb{R}^N$ is the latent state, and $\mathbf{y}(t) \in \mathbb{R}^O$ is the output. The matrices $\mathbf{A}, \mathbf{B}, \mathbf{C}, \mathbf{D}$ govern the evolution and output generation mechanisms. Early work on SSMs explored constrained parameterizations, such as diagonal initialization of $\mathbf{A}$ \cite{gu2021efficiently}, which offered simple and effective control over temporal dynamics.
More advanced and recent formulations \cite{fuhungry, peng2023rwkv, poli2023hyena} introduced input-dependent and learnable SSM parameters, enabling models to selectively amplify relevant information or attenuate irrelevant patterns within the sequence. This concept was further refined in the popular Mamba architecture \cite{gu2023mamba}. In this design, the Selective SSM forms the core of a new modular building block, the Mamba Block. This enables linear-time processing with respect to sequence length, offering a powerful alternative to the quadratic cost of attention-based models.
Encouraged by this success, Mamba Blocks have also been used for vision detection tasks, as demonstrated in \cite{zhu2024vision, hatamizadeh2024mambavision}, where sequences were formed by scanning flattened image patches from multiple directions. Moreover, Mamba has also been successfully applied in the hyperspectral imaging domain, such as for classification \cite{yao2024spectralmamba, huang2024spectral} and detection \cite{shen2025htd}.

\begin{table}[t]
\caption{HySpecNet-11k dataset. $4\times$ super-resolution. Parameters and FLOPs/pixel for input size $32 \times 32 \times 202$.}
\label{tab:hysp_x4}
\setlength{\tabcolsep}{0.3mm}
\centering
\begin{tabular}{ccccccccc}
\textbf{Method}  & \textbf{MPSNR(dB)}$\uparrow$ & \textbf{MSSIM}$\uparrow$ & \textbf{SAM}$\downarrow$ & \textbf{RMSE}$\downarrow$ & \textbf{Params} & \textbf{FLOPs/px} \\
\hline
\hline
{Bicubic}  & 41.00 & 0.9004 &  4.1767 & 0.0147 & -- & --  \\ 
{GDRRN \cite{li2018single}} & 41.42 & 0.9167 & 4.0681 & 0.0132  & 613 K &  284 K  \\ 
{SSPSR \cite{jiang2020learning}} & 42.90 & 0.9413 & 3.6765 &  0.0108    & 14.07 M  & 2569 K  \\
{SNLSR \cite{hu2024exploring}} &  42.94 & 0.9389 & 3.6062 &    0.0110  & 1.81 M  & 35 K \\
{EUNet \cite{liu2023efficient}} & 42.96 &  0.9386 & 3.6011 & 0.0110    & 1.01 M & 186 K  \\
{MSDformer \cite{chen2023msdformer}} & 43.45 & 0.9469 &  3.3702 &  0.0103   & 17.49 M & 714 K \\
{CST \cite{chen2024cross}} & 43.61 & 0.9495 & 3.3599 & 0.0100 &  11.48 M & 245 K \\
{\textbf{DPSR}} & 43.17  & 0.9432 & 3.5010 & 0.0106 & 2.71 M & 31 K \\
\bottomrule
\end{tabular}
\end{table}

\begin{table}
\caption{HySpecNet-11k dataset. $2\times$ super-resolution.}
\label{tab:hysp_x2}
\setlength{\tabcolsep}{0.3mm}
\centering
\begin{tabular}{ccccccccccc}
\textbf{Method}  & \textbf{MPSNR(dB)}$\uparrow$ & \textbf{MSSIM}$\uparrow$ & \textbf{SAM}$\downarrow$ & \textbf{RMSE}$\downarrow$ & \textbf{Params} & \textbf{FLOPs/px}  \\
\hline
\hline
{Bicubic}  & 45.68 & 0.9654 & 2.8981   & 0.0083 & -- & -- \\ 
{GDRRN \cite{li2018single}} & 47.30 & 0.9802 & 2.6726 & 0.0063 & 613 K & 71 K \\ 
{SSPSR \cite{jiang2020learning}} & 48.56 & 0.9859 & 2.3735 & 0.0053 & 11.72 M & 1430 K \\
{SNLSR \cite{hu2024exploring}} & -- & -- & -- & -- & -- & -- \\
{EUNet \cite{liu2023efficient}} & 48.62 & 0.9850  & 2.3411 & 0.0055  & 944 K & 65 K \\
{MSDformer \cite{chen2023msdformer}} & 49.44 & 0.9879 & 2.0866 & 0.0049 & 15.42 M & 528 K   \\
{CST \cite{chen2024cross}} & 49.77 & 0.9894 & 2.0900 & 0.0045 & 10.31 M & 121 K   \\
{\textbf{DPSR}} & 48.85 & 0.9862  &   2.2644  & 0.0052  & 2.07 M & 20 K \\
\bottomrule
\end{tabular}
\end{table}

\begin{table}[t]
\caption{Houston dataset. $4\times$ super-resolution.}
\label{tab:houston_x4}
\setlength{\tabcolsep}{0.7mm}
\centering
\begin{tabular}{ccccccccc}
\textbf{Method}  & \textbf{MPSNR(dB)}$\uparrow$ & \textbf{MSSIM}$\uparrow$ & \textbf{SAM}$\downarrow$ & \textbf{RMSE}$\downarrow$    \\
\hline
\hline
{Bicubic}  & 44.21 & 0.9657 & 2.4038 & 0.0074 \\ 
{GDRRN \cite{li2018single}}  & 45.80 & 0.9772 & 2.0724 & 0.0059 \\ 
{SSPSR \cite{jiang2020learning}}  & 46.93 & 0.9814 & 1.7369 & 0.0053 \\
{SNLSR \cite{hu2024exploring}} & 46.84 & 0.9808 & 1.7581 & 0.0054   \\
{EUNet \cite{liu2023efficient}} & 46.81 & 0.9808 & 1.7227 & 0.0054   \\
{MSDformer \cite{chen2023msdformer}}  & 46.85 & 0.9808 & 1.7244 & 0.0054  \\
{CST \cite{chen2024cross}} & 47.63 &  0.9840 & 1.6358 & 0.0049    \\
{\textbf{DPSR}} & 47.53 & 0.9834  & 1.6451 & 0.0050     \\
\bottomrule
\end{tabular}
\end{table}

\begin{table}[t]
\caption{Houston dataset. $2\times$ super-resolution.}
\label{tab:houston_x2}
\setlength{\tabcolsep}{0.7mm}
\centering
\begin{tabular}{ccccccccc}
\textbf{Method}  & \textbf{MPSNR(dB)}$\uparrow$ & \textbf{MSSIM}$\uparrow$ & \textbf{SAM}$\downarrow$ & \textbf{RMSE}$\downarrow$   \\
\hline
\hline
{Bicubic}  & 50.75 & 0.9927 & 1.2085 & 0.0034 \\ 
{GDRRN \cite{li2018single}}  &  52.68 & 0.9954 & 1.0355 & 0.0027   \\ 
{SSPSR \cite{jiang2020learning}} &  53.24 & 0.9959 & 0.9531 & 0.0025   \\
{SNLSR \cite{hu2024exploring}}  & -- & --  & --  & --   \\
{EUNet \cite{liu2023efficient}}  &  53.51 & 0.9961 & 0.9089 & 0.0024   \\
{MSDformer \cite{chen2023msdformer}} & 53.25 &0.9959&0.9437&0.0025    \\
{CST \cite{chen2024cross}}  &  53.84 & 0.9964 & 0.8866 & 0.0023  \\
{\textbf{DPSR}} & 53.57 & 0.9961 & 0.9103 & 0.0024     \\
\bottomrule
\end{tabular}
\end{table}

In our setting, the model must process a continuous stream of lines in the along-track direction with causal propagation of the information and a limited memory and computational budget. Mamba is the most suitable approach compared to attention-based or recurrent methods. Attention-based methods scale quadratically with sequence length, whereas Mamba scales linearly thus limiting complexity. Recurrent designs are causal but they have limited effective context, and do not allow efficient training while Mamba's selective state updates provide a better way to store and retrieve relevant past information.

\subsection{Onboard AI} \label{sec:onboard_ai}
Over the past few years, onboard AI-based processing has emerged as a powerful method for performing real-time analysis of satellite data, significantly reducing latency for tasks requiring immediate response and optimizing bandwidth usage. The ability to process large volumes of imagery directly in space can indeed enable more responsive satellite operations, facilitate prompt detection of critical events, and open the door to autonomous decision-making. Nonetheless, the design of such onboard AI solutions faces multiple challenges, including strict power constraints, limited onboard memory, and the need for robust, lightweight architectures that can function reliably in harsh conditions. 

Early works proved the concept on narrowly scoped problems. In \cite{yao2019board} the authors squeezed a pruned YOLO variant onto a CubeSat for ship spotting, and in \cite{giuffrida2021varphi} a compact cloud-mask CNN was launched on ESA's $\Phi$-Sat-1 to perform cloud-detection. Those successes triggered broader surveys: in \cite{ziaja2021benchmarking} many networks were benchmarked against flight-qualified FPGAs and CPUs; in \cite{ruuvzivcka2022ravaen} distillation was used to produce a lean variational auto-encoder for change detection. In \cite{kervennic2023embedded} the authors report three years of OPS-SAT trials in which an embedded U-Net performs pixel-level cloud segmentation. More recently, a new model was proposed in  \cite{inzerillo2024efficient} to perform onboard multitask inference. Looking toward constellation scenarios, in \cite{qiao2024orbit} the authors tried distributing a deep network across multiple platforms, securing a $40\%$ energy saving. More similarly to the work in this paper, a line processing concept for onboard image compression was designed in \cite{valsesia2024onboard}. However, they use a RWKV hybrid attention-recurrent mechanism and their framework relies on a pixel-by-pixel predictive coding approach.
Collectively, these efforts show that while there are still hard computational and memory constraints, onboard AI is turning from a curiosity into a mission enabler, edging the satellite community ever closer to fully autonomous operations.

\section{Proposed Method}

In this section, we introduce the method that we propose to address the problem of onboard single-image super-resolution and the structure of our line-by-line neural network architecture, called DPSR. A high-level illustration of the concept is given in Figure \ref{fig:method}, while a detail of the DPSR neural network architecture is shown in Figure \ref{fig:architecture}. 

\subsection{Preliminaries and Proposed Strategy}
The overall design goal is to super-resolve an image line-by-line in the along-track direction, to match a pushbroom acquisition system and minimize memory and computational requirements. In the remainder of the paper, we will use $H$ to denote the number of lines in the along-track direction $y$, $W$ to denote the number of columns in the across-track direction $x$ and $C$ to denote the number of spectral channels.
We also remark that super-resolving a line, with all its spectral channels, of size $1 \times W \times C$ by a factor $r$ produces an output of size $r \times rW \times C$, i.e., $r$ lines in the along-track direction, expanded by a factor of $r$ in the across-track direction.

More formally, we suppose to have acquired the current LR line $\mathbf{x}^\text{LR}_y \in \mathbb{R}^{W \times C}$ corresponding to along-track LR line location $y$ and a memory of past lines which will be described in Sec. \ref{sec:arch}. 
From this, we estimate $\lbrace \mathbf{x}^\text{SR}_{r(y-1)-(r-1)}, \mathbf{x}^\text{SR}_{r(y-1)-(r-2)}, ... , \mathbf{x}^\text{SR}_{r(y-1)-1} , \mathbf{x}^\text{SR}_{r(y-1)}  \rbrace$ with $y = 1, \dots$, i.e., the set of $r$ super-resolved (SR) lines at the high-resolution (HR) along-track spatial locations between the LR spatial locations $y-1$ (included) and $y$ (excluded). It is important to remark that we are framing the problem with an \textit{interpolation} setting, where line $y$ is used to predict HR locations strictly preceding it. This is in contrast with an \textit{extrapolation} setting which would predict spatial locations beyond $y$, and would be more challenging leading to worse image quality.
Moreover, the estimation of the SR lines is performed in a residual fashion, i.e., a neural network only estimates an additive correction to the result of a bilinear interpolator.

We note that this configuration restricts the model to using only previously acquired lines, with no access to future lines. This departs from existing methods, which require a 2D tile of the image. This choice is intentional since such methods would require to buffer a large number of lines to form a complete image and their processing would require significantly increased memory and computational resources. Essentially, DPSR trades access to future lines, which may potentially limit quality, for complexity and speed.

\begin{table}[t]
\caption{Pavia dataset. $4\times$ super-resolution.}
\label{tab:pavia_x4}
\setlength{\tabcolsep}{0.7mm}
\centering
\begin{tabular}{ccccccccc}
\textbf{Method}  & \textbf{MPSNR(dB)}$\uparrow$ & \textbf{MSSIM}$\uparrow$ & \textbf{SAM}$\downarrow$ & \textbf{RMSE}$\downarrow$   \\
\hline
\hline
{Bicubic}  & 27.69 & 0.7332 & 6.5587 & 0.0429 \\ 
{GDRRN \cite{li2018single}}  & 28.70 & 0.7981 & 6.4347 & 0.0378 \\ 
{SSPSR \cite{jiang2020learning}}  & 28.99 & 0.8116 & 5.7788 & 0.0367 \\
{SNLSR \cite{hu2024exploring}} & 28.78 &  0.8038 & 5.8878 &  0.0374     \\
{EUNet \cite{liu2023efficient}} & 28.80 & 0.8037 & 5.7460 & 0.0374   \\
{MSDformer \cite{chen2023msdformer}}  &  28.95 & 0.8088 & 5.7940 & 0.0368 \\
{CST \cite{chen2024cross}} & 29.13 & 0.8163 & 5.7399 & 0.0360   \\
{\textbf{DPSR}} & 29.05 & 0.8151 & 5.9737 &  0.0363     \\
\bottomrule
\end{tabular}
\end{table}

\begin{table}[t]
\caption{Pavia dataset. $2\times$ super-resolution.}
\label{tab:pavia_x2}
\setlength{\tabcolsep}{0.7mm}
\centering
\begin{tabular}{ccccccccc}
\textbf{Method}  & \textbf{MPSNR(dB)}$\uparrow$ & \textbf{MSSIM}$\uparrow$ & \textbf{SAM}$\downarrow$ & \textbf{RMSE}$\downarrow$   \\
\hline
\hline
{Bicubic}  & 32.39 & 0.9187 & 4.6856 & 0.0245  \\ 
{GDRRN \cite{li2018single}}  & 34.89 & 0.9513 & 4.0743 & 0.0183   \\ 
{SSPSR \cite{jiang2020learning}} & 35.73 & 0.9598 & 3.6963 & 0.0166   \\
{SNLSR \cite{hu2024exploring}}  & -- & --  & --  & --   \\
{EUNet \cite{liu2023efficient}}  & 35.80 & 0.9604 & 3.6056 & 0.0165   \\
{MSDformer \cite{chen2023msdformer}} & 35.86 & 0.9610 & 3.6323 & 0.0164   \\
{CST \cite{chen2024cross}}  & 36.17 & 0.9634 & 3.5850 & 0.0158 \\
{\textbf{DPSR}} &  35.87 & 0.9612 & 3.6160 &   0.0164  \\
\bottomrule
\end{tabular}
\end{table}

\subsection{DPSR architecture}
\label{sec:arch}

\begin{table}[t]
\caption{Chikusei dataset. $4\times$ super-resolution.}
\label{tab:chiku_x4}
\setlength{\tabcolsep}{0.7mm}
\centering
\begin{tabular}{ccccccccc}
\textbf{Method}  & \textbf{MPSNR(dB)}$\uparrow$ & \textbf{MSSIM}$\uparrow$ & \textbf{SAM}$\downarrow$ & \textbf{RMSE}$\downarrow$ \\
\hline
\hline
{Bicubic} & 37.64 & 0.8954 & 3.4040 & 0.0156    \\ 
{GDRRN \cite{li2018single}} & 39.09 & 0.9265 & 3.0536 & 0.0130   \\ 
{SSPSR \cite{jiang2020learning}} & 39.98  & 0.9393 & 2.4864 & 0.0119  \\
{SNLSR \cite{hu2024exploring}} & 39.90 &  0.9387 & 2.4722 & 0.0119 &   \\
{EUNet \cite{liu2023efficient}} & 39.94 &   0.9398  & 2.3923 & 0.0121  \\
{MSDformer \cite{chen2023msdformer}} & 40.09 & 0.9405 & 2.3981 & 0.0118   \\
{CST \cite{chen2024cross}} & 40.24 & 0.9431 & 2.3453 & 0.0116  \\
{\textbf{DPSR}} & 40.07 & 0.9410 &  2.3695  &   0.0118  \\
\bottomrule
\end{tabular}
\end{table}

\begin{table}[t]
\caption{Chikusei dataset. $2\times$ super-resolution.}
\label{tab:chiku_x2}
\setlength{\tabcolsep}{0.7mm}
\centering
\begin{tabular}{ccccccccc}
\textbf{Method}  & \textbf{MPSNR(dB)}$\uparrow$ & \textbf{MSSIM}$\uparrow$ & \textbf{SAM}$\downarrow$ & \textbf{RMSE}$\downarrow$ \\
\hline
\hline
{Bicubic}  & 43.21 & 0.9721  & 1.7880 & 0.0082   \\ 
{GDRRN \cite{li2018single}} & 46.43 & 0.9869 & 1.3911  & 0.0056  \\ 
{SSPSR \cite{jiang2020learning}} & 47.41 & 0.9893 & 1.2035 & 0.0051  \\
{SNLSR \cite{hu2024exploring}} & -- & -- & -- & --   \\
{EUNet \cite{liu2023efficient}} & 47.82 & 0.9903 & 1.1143 & 0.0048     \\
{MSDformer \cite{chen2023msdformer}} & 47.69 & 0.9899 & 1.1457 & 0.0049   \\
{CST \cite{chen2024cross}} & 48.09 &  0.9908 & 1.1057   & 0.0047     \\
{\textbf{DPSR}} & 47.49 & 0.9894 & 1.1461 & 0.0050  \\
\bottomrule
\end{tabular}
\end{table}

This section introduces the neural network architecture of DPSR, which is shown in Figs. \ref{fig:architecture}, \ref{fig:SFE+Up}.

As a first step, we decouple the along-track dimension from the across-track dimension by extracting the line $\mathbf{x}_{y}^{LR}$ along with all its spectral channels. This leads to the input of the architecture having size of $1 \times W \times C$. Conceptually, DPSR will first extract features that jointly represent the pixels in the across-track and spectral dimensions of the current line, and then use a Selective SSM (Mamba \cite{gu2023mamba}) to integrate these features with a memory of those of previous lines.

This results in a design featuring an initial Shallow Feature Extraction Block (SFE Block) comprising 1D operations operating on across-track pixels and spectral channels. In particular, a 1D convolution layer, with a spatial kernel size of 3, transforms the input channels $C$ into an arbitrary number of features $F$. This is then followed by layer normalization and nonlinear activation, for which we used the Sigmoid Linear Unit (SiLU) function \cite{hendrycks2016gaussian}. Subsequently, a Channel Attention Block \cite{woo2018cbam} is introduced. We denote the output of the SFE block as $\mathbf{z}_y \in \mathbb{R}^{W \times F}$ where each pixel in the across-track direction is associated to a feature vector jointly capturing correlations between the columns and the original spectral bands.
The SFE Block is then followed by a backbone composed of two Cross-Line Feature Fusion (CLFF) Blocks, each consisting of a NAFBlock followed by a Mamba Block. The design of the CLFF Blocks is motivated by the different but complementary roles of the NAFBlock and Mamba Block. The tensor $\mathbf{z}_{y}$ obtained as output of the SFE Block is fed to the CLFF Blocks: 

\begin{align}
    \mathbf{w}^{(2)}_y &= \text{CLFF}_{2}\!\bigl(\text{CLFF}_{1}(\mathbf{z}_y)\bigr).
\end{align}

More in detail, the NAFBlock uses a sequence of layers inspired by the building blocks of the NAFNet architecture \cite{chen2022simple} to expand the across-track receptive field and extract deeper features in the across-track and spectral dimensions. In particular, this block uses separable convolutions (size-1 1D convolution, followed by depthwise 1D convolution) to achieve a lightweight receptive field expansion. Moreover, a SimpleGate and a Simplified Channel Attention operations implement an attention-like mechanism of input-dependent feature extraction, while keeping a lightweight design employing no specific activation function. We refer the reader to \cite{chen2022simple} for more details on these operations.

\captionsetup[subfigure]{%
  font=small,
  labelfont={small},
  justification=centering,
  position=top,
  skip=1pt              
}

\begin{figure}[t]
\centering
\begin{subfigure}[b]{0.32\linewidth}
  \caption{\small Bicubic}
  \includegraphics[width=\linewidth]{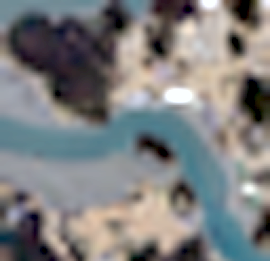}
\end{subfigure}
\begin{subfigure}[b]{0.32\linewidth}
  \caption{\small GDRRN~\cite{li2018single}}
  \includegraphics[width=\linewidth]{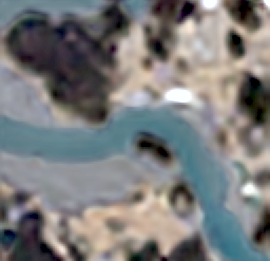}
\end{subfigure}
\begin{subfigure}[b]{0.32\linewidth}
  \caption{\small SSPSR~\cite{jiang2020learning}}
  \includegraphics[width=\linewidth]{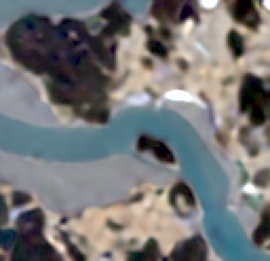}
\end{subfigure}

\vspace{4pt} 
\begin{subfigure}[b]{0.32\linewidth}
  \caption{\small SNLSR~\cite{hu2024exploring}}
  \includegraphics[width=\linewidth]{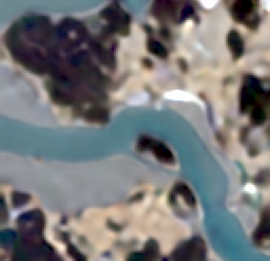}
\end{subfigure}
\begin{subfigure}[b]{0.32\linewidth}
  \caption{\small EUNet~\cite{liu2023efficient}}
  \includegraphics[width=\linewidth]{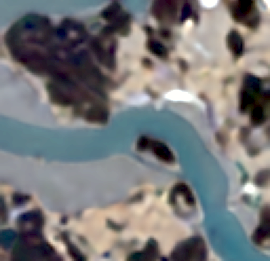}
\end{subfigure}
\begin{subfigure}[b]{0.32\linewidth}
  \caption{\small MSDformer~\cite{chen2023msdformer}}
  \includegraphics[width=\linewidth]{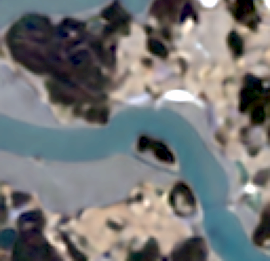}
\end{subfigure}
\vspace{4pt} 
\begin{subfigure}[b]{0.32\linewidth}
  \caption{\small CST~\cite{chen2024cross}}
  \includegraphics[width=\linewidth]{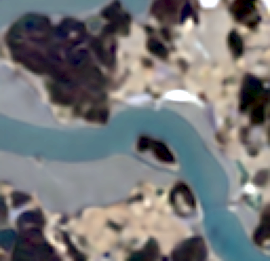}
\end{subfigure}
\begin{subfigure}[b]{0.32\linewidth}
  \caption{\small DPSR}
  \includegraphics[width=\linewidth]{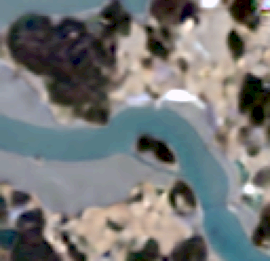}
\end{subfigure}
\begin{subfigure}[b]{0.32\linewidth}
  \caption{\small Ground Truth}
  \includegraphics[width=\linewidth]{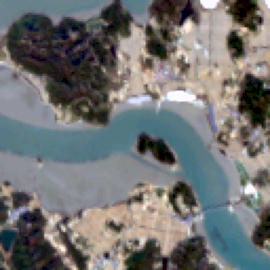}
\end{subfigure}

\caption{Qualitative comparison of methods on HySpecNet-11k test image 247 with SR factor $4\times$. Bands 43, 28, 10 were used as R, G, B.}
\label{fig:test_247}
\end{figure}

\begin{figure}[t]
\centering
\begin{subfigure}[b]{0.32\linewidth}
  \caption{\small Bicubic}
  \includegraphics[width=\linewidth]{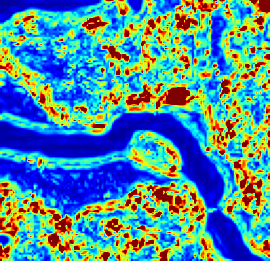}
\end{subfigure}
\begin{subfigure}[b]{0.32\linewidth}
  \caption{\small GDRRN~\cite{li2018single}}
  \includegraphics[width=\linewidth]{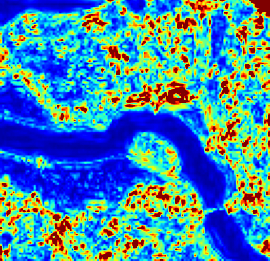}
\end{subfigure}
\begin{subfigure}[b]{0.32\linewidth}
  \caption{\small SSPSR~\cite{jiang2020learning}}
  \includegraphics[width=\linewidth]{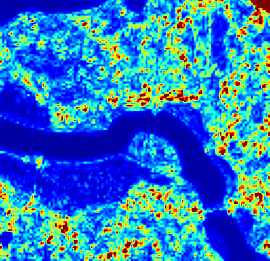}
\end{subfigure}

\vspace{4pt}
\begin{subfigure}[b]{0.32\linewidth}
  \caption{\small SNLSR~\cite{hu2024exploring}}
  \includegraphics[width=\linewidth]{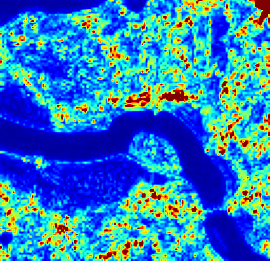}
\end{subfigure}
\begin{subfigure}[b]{0.32\linewidth}
  \caption{\small EUNet~\cite{liu2023efficient}}
  \includegraphics[width=\linewidth]{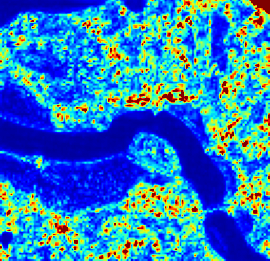}
\end{subfigure}
\begin{subfigure}[b]{0.32\linewidth}
  \caption{\small MSDformer~\cite{chen2023msdformer}}
  \includegraphics[width=\linewidth]{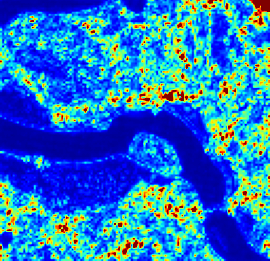}
\end{subfigure}

\vspace{4pt}
\begin{subfigure}[b]{0.32\linewidth}
  \caption{\small CST~\cite{chen2024cross}}
  \includegraphics[width=\linewidth]{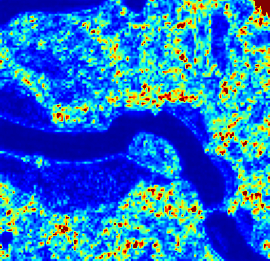}
\end{subfigure}
\begin{subfigure}[b]{0.32\linewidth}
  \caption{\small DPSR}
  \includegraphics[width=\linewidth]{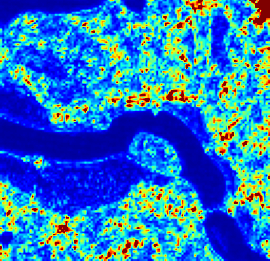}
\end{subfigure}
\begin{subfigure}[b]{0.32\linewidth}
  \mbox{} 
\end{subfigure}

\caption{Mean (across bands) error heatmaps of methods on HySpecNet-11k test image 247 with SR factor $4\times$.}
\label{fig:test_247_error}
\end{figure}

\begin{figure}[t]
\centering
\begin{subfigure}[b]{0.32\linewidth}
  \caption{\small Bicubic}
  \includegraphics[width=\linewidth]{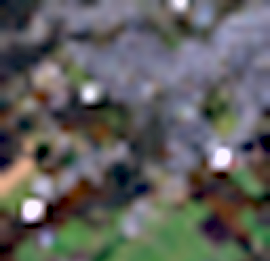}
\end{subfigure}
\begin{subfigure}[b]{0.32\linewidth}
  \caption{\small GDRRN~\cite{li2018single}}
  \includegraphics[width=\linewidth]{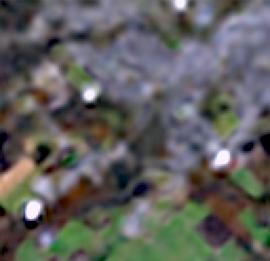}
\end{subfigure}
\begin{subfigure}[b]{0.32\linewidth}
  \caption{\small SSPSR~\cite{jiang2020learning}}
  \includegraphics[width=\linewidth]{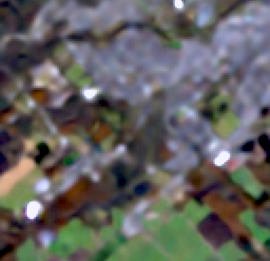}
\end{subfigure}

\vspace{4pt}
\begin{subfigure}[b]{0.32\linewidth}
  \caption{\small SNLSR~\cite{hu2024exploring}}
  \includegraphics[width=\linewidth]{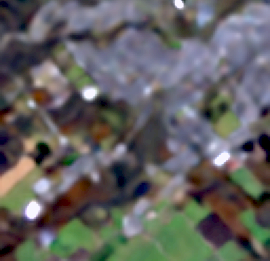}
\end{subfigure}
\begin{subfigure}[b]{0.32\linewidth}
  \caption{\small EUNet~\cite{liu2023efficient}}
  \includegraphics[width=\linewidth]{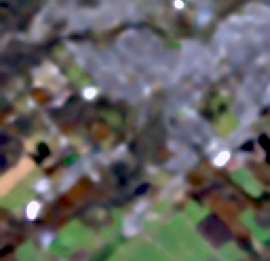}
\end{subfigure}
\begin{subfigure}[b]{0.32\linewidth}
  \caption{\small MSDformer~\cite{chen2023msdformer}}
  \includegraphics[width=\linewidth]{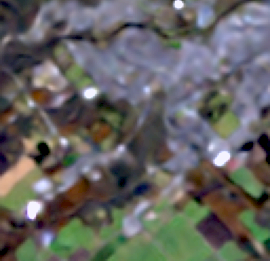}
\end{subfigure}

\vspace{4pt}
\begin{subfigure}[b]{0.32\linewidth}
  \caption{\small CST~\cite{chen2024cross}}
  \includegraphics[width=\linewidth]{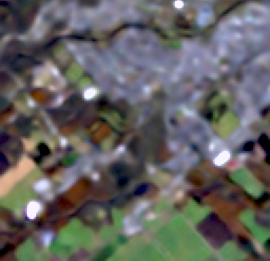}
\end{subfigure}
\begin{subfigure}[b]{0.32\linewidth}
  \caption{\small DPSR}
  \includegraphics[width=\linewidth]{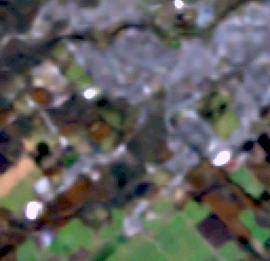}
\end{subfigure}
\begin{subfigure}[b]{0.32\linewidth}
  \caption{\small Ground Truth}
  \includegraphics[width=\linewidth]{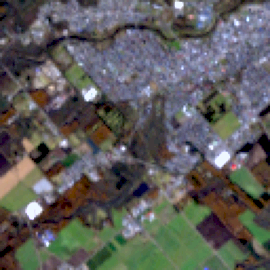}
\end{subfigure}

\caption{Qualitative comparison of methods on HySpecNet-11k test image 628 with SR factor $4\times$. Bands 43, 28, 10 were used as R, G, B.}
\label{fig:test_628}
\end{figure}

\begin{figure}[t]
\centering
\begin{subfigure}[b]{0.32\linewidth}
  \caption{\small Bicubic}
  \includegraphics[width=\linewidth]{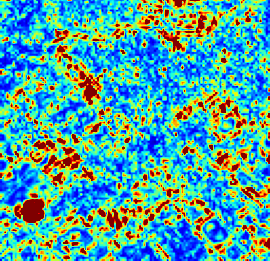}
\end{subfigure}
\begin{subfigure}[b]{0.32\linewidth}
  \caption{\small GDRRN~\cite{li2018single}}
  \includegraphics[width=\linewidth]{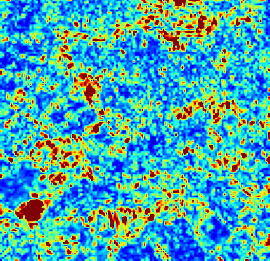}
\end{subfigure}
\begin{subfigure}[b]{0.32\linewidth}
  \caption{\small SSPSR~\cite{jiang2020learning}}
  \includegraphics[width=\linewidth]{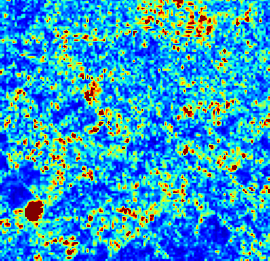}
\end{subfigure}

\vspace{4pt}
\begin{subfigure}[b]{0.32\linewidth}
  \caption{\small SNLSR~\cite{hu2024exploring}}
  \includegraphics[width=\linewidth]{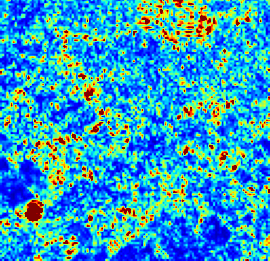}
\end{subfigure}
\begin{subfigure}[b]{0.32\linewidth}
  \caption{\small EUNet~\cite{liu2023efficient}}
  \includegraphics[width=\linewidth]{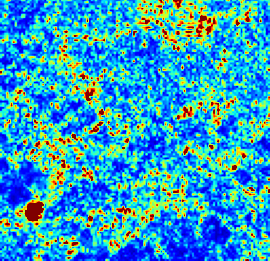}
\end{subfigure}
\begin{subfigure}[b]{0.32\linewidth}
  \caption{\small MSDformer~\cite{chen2023msdformer}}
  \includegraphics[width=\linewidth]{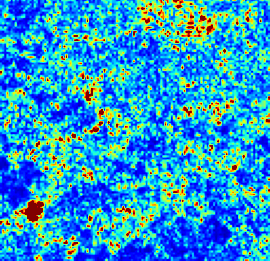}
\end{subfigure}

\vspace{4pt}
\begin{subfigure}[b]{0.32\linewidth}
  \caption{\small CST~\cite{chen2024cross}}
  \includegraphics[width=\linewidth]{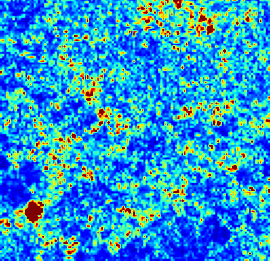}
\end{subfigure}
\begin{subfigure}[b]{0.32\linewidth}
  \caption{\small DPSR}
  \includegraphics[width=\linewidth]{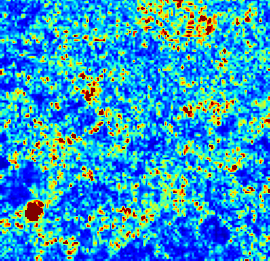}
\end{subfigure}
\begin{subfigure}[b]{0.32\linewidth}
  \mbox{} 
\end{subfigure}

\caption{Mean (across bands) error heatmaps of methods on HySpecNet-11k test image 628 with SR factor $4\times$.}
\label{fig:test_628_error}
\end{figure}

\begin{figure}[t]
\centering
\begin{subfigure}[b]{0.32\linewidth}
  \caption{\small Bicubic}
  \includegraphics[width=\linewidth]{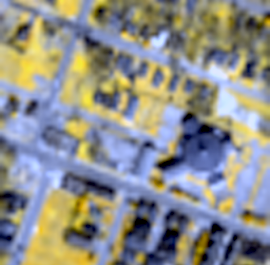}
\end{subfigure}
\begin{subfigure}[b]{0.32\linewidth}
  \caption{\small GDRRN~\cite{li2018single}}
  \includegraphics[width=\linewidth]{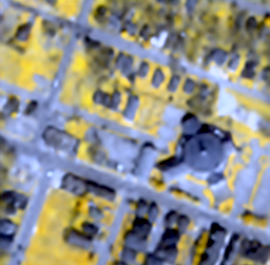}
\end{subfigure}
\begin{subfigure}[b]{0.32\linewidth}
  \caption{\small SSPSR~\cite{jiang2020learning}}
  \includegraphics[width=\linewidth]{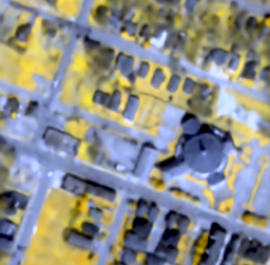}
\end{subfigure}

\vspace{4pt}
\begin{subfigure}[b]{0.32\linewidth}
  \caption{\small SNLSR~\cite{hu2024exploring}}
  \includegraphics[width=\linewidth]{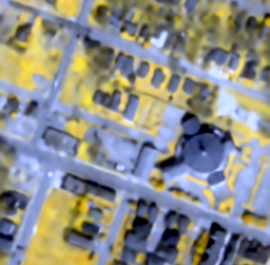}
\end{subfigure}
\begin{subfigure}[b]{0.32\linewidth}
  \caption{\small EUNet~\cite{liu2023efficient}}
  \includegraphics[width=\linewidth]{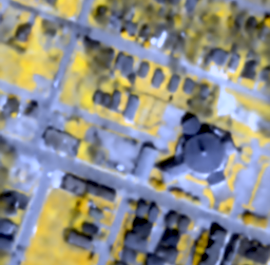}
\end{subfigure}
\begin{subfigure}[b]{0.32\linewidth}
  \caption{\small MSDformer~\cite{chen2023msdformer}}
  \includegraphics[width=\linewidth]{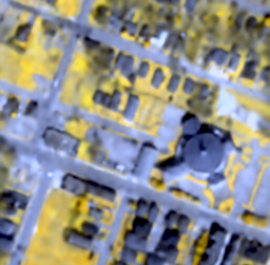}
\end{subfigure}

\vspace{4pt}
\begin{subfigure}[b]{0.32\linewidth}
  \caption{\small CST~\cite{chen2024cross}}
  \includegraphics[width=\linewidth]{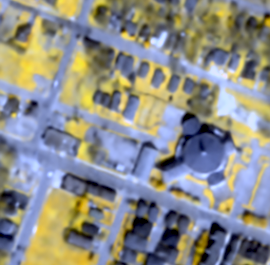}
\end{subfigure}
\begin{subfigure}[b]{0.32\linewidth}
  \caption{\small DPSR}
  \includegraphics[width=\linewidth]{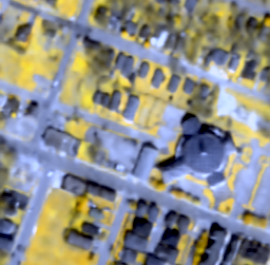}
\end{subfigure}
\begin{subfigure}[b]{0.32\linewidth}
  \caption{\small Ground Truth}
  \includegraphics[width=\linewidth]{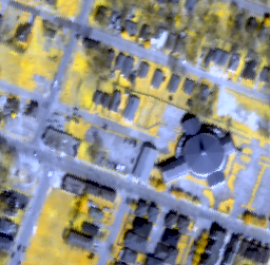}
\end{subfigure}

\caption{Qualitative comparison of methods on a Houston test image with SR factor $4\times$. Bands 29, 26, 19 were used as R, G, B following \cite{chen2023msdformer}.}
\label{fig:image_houston}
\end{figure}

\begin{figure}
    \centering
    \includegraphics[width=0.47\textwidth]{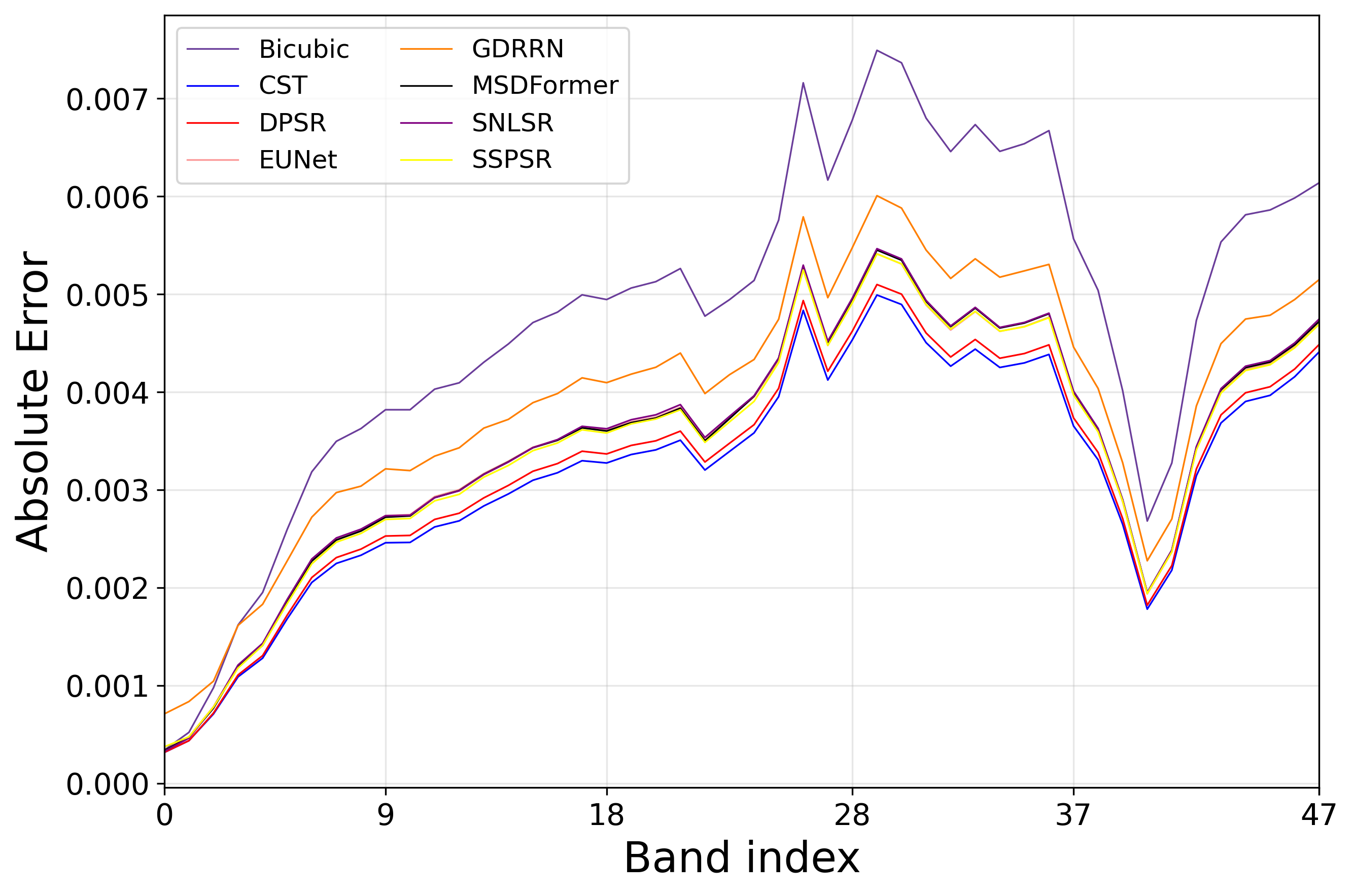}
    \caption{Per-band mean absolute error across the SR images of the Houston test set.}
    \label{fig:absolute_error}
\end{figure}

\begin{figure*}
    \centering
    \includegraphics[width=1.\textwidth]{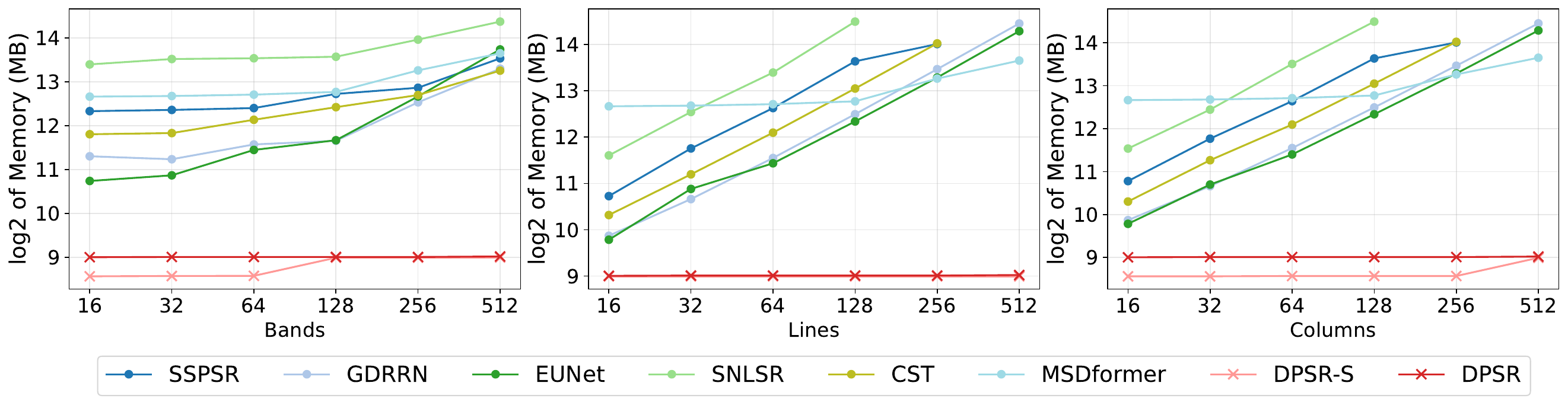}
    \caption{Memory usage comparison for $4\times$ SR. Input image size is a) $256 \times 256 \times C$, b) $H \times 1000 \times 66$, c) $1000 \times W \times 66$. SNLSR runs out of memory ($>24$GB) when $H,W$ exceed 128 pixels, and CST and SSPSR when $H,W$ exceed 256 pixels. Note that CST does not accept spatial size of $1000$ as input, so we set it to $992$. }
    \label{fig:memory_dims_x4}
\end{figure*}

\begin{figure*}
    \centering
    \includegraphics[width=1.\textwidth]{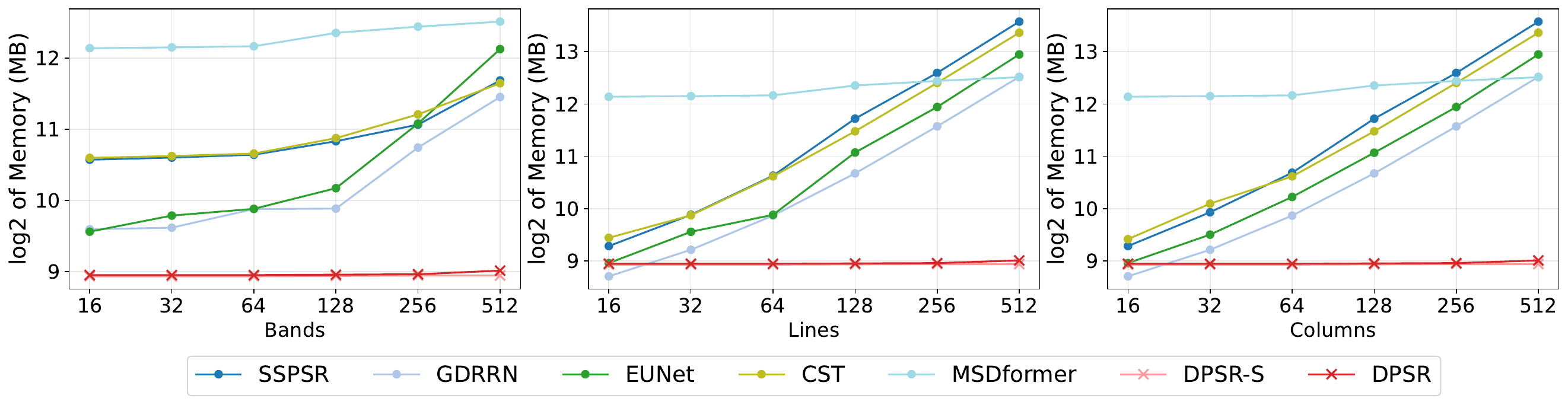}
    \caption{Memory usage comparison for $2\times$ SR. Input image size is a) $256 \times 256 \times C$, b) $H \times 1000 \times 66$, c) $1000 \times W \times 66$. Note that CST does not accept spatial size of $1000$ as input, so we set it to $992$. SNLSR does not support $2\times$ SR.}
    \label{fig:memory_dims_x2}
\end{figure*}

On the other hand, the Mamba Block \cite{gu2023mamba} has the crucial job of learning the interactions between successive lines, i.e., the propagation of information in the along-track dimension. The output of the NAFBlock consists of one feature vector for each of the $W$ pixels in the line, which can be regarded as a token whose evolution is modeled over the along-track direction by means of Mamba.  In the Mamba Block, the input feature dimension $F$ is expanded to $EF$, where $E$ is an arbitrary constant. Then, a causal convolution operation in the along-track direction further processes the feature sequence. Notice that this requires storing the receptive of the causal convolution operator, i.e., the features of $K$ past lines, where $K$ is the convolution kernel size. We remark that $K$ is typically a small value, such as $K=4$ as we used in the experiments, so that a very modest number of past lines needs to be retained. After non-linear activation with a SiLU function, the line features $\mathbf{z}'_y \in \mathbb{R}^{W \times EF}$ are processed by the Mamba Selective SSM formulation introduced in Sec. \ref{sec:sequencemodeling} so that long-term dependencies encoded in the latent state can be exploited. The latent state of the SSM is a vector of length $N$ for each expanded feature of the line pixels, i.e., $\mathbf{h} \in \mathbb{R}^{W \times EF \times N}$ and can be seen as a condensed running memory of all the previous lines scanned by the model, which allows the model to uncover self-similarities in the far past, dropping the requirement of a buffer of a high number of lines in order to perform effective super-resolution. In formulas,  when the SiLU-processed features $\mathbf{z'}_y$ reach the state-space module, they are processed by a discrete-time version of the linear SSM in Sec. \ref{sec:sequencemodeling}:       
\begin{align}
\mathbf{h}_{y} &= \mathbf{A}_{y}\mathbf{h}_{y-1} + \mathbf{B}_{y}\mathbf{z'}_{y}, \label{eq:memory_update} \\
\mathbf{z''}_{y} &= \mathbf{C}_{y}\mathbf{h}_{y} + \mathbf{D}_{y}\mathbf{z'}_{y}, \label{eq:features_update}
\end{align}
operating in parallel over all the $W$ pixels in the current line. The state $\mathbf{h}$ is updated at each time step, i.e. every time a new line is processed, following Eq. \eqref{eq:memory_update}. New context-rich line features are then produced as output of Eq. \eqref{eq:features_update} using the updated latent state. Finally, the SSM-transformed features are gated by the original features input to the Mamba Block after feature expansion and SiLU non-linearity, and projected back to $F$-dimensional space.
The Mamba Block is the main driver of memory requirements for the architecture and, based on the aforementioned design, we can summarize such requirements as: i) a tensor of size $K \times W \times EF$ with the features of current line and past lines in the receptive field of causal convolution; ii) a tensor of size $W \times EF \times N$ with the latent state of the SSM. These are quite modest memory requirements: for instance, if we consider a realistic $W=1000$, $F=128$, $E=1$, $N=16$ setting, then only about 10 MB of memory is required to store the aforementioned tensors as single-precision floating point values.

The final part of the architecture is the upsampler module, which is responsible for the output tensor to have the right spatial dimensions, according to the desired super-resolution scale factor. The upsampler first expands the input tensor channels with a 1D convolution to a value $fr^2$ where $r$ is the SR factor and $f$ an arbitrary number of features that will remain after the subsequent Pixel Shuffle \cite{shi2016real}. The last layer is a 1D convolution that restores the number of channels of the input image. We remark that choosing a suitable value for $f$ is tradeoff between limiting the amount of floating point operations performed by the upsampler module and properly representing all the spectral channels. Indeed, it could be argued that the optimal value $f$ is tied to how much the spectral channels are redundant and thus their dimensionality could be reduced. Hence, the final super-resolved lines are obtained as:
\begin{align}
    &\mathbf{x}^\text{SR}_{r(y-1)}, ..., \mathbf{x}^\text{SR}_{r(y-1)-(r-1)} =\\
    &=\text{Upsampler}(\mathbf{w}^{(2)}_y) + \text{Bilinear}(\mathbf{x}^\text{LR}_y,\mathbf{x}^\text{LR}_{y-1}).
\end{align}

\section{Experimental results}

In this section we analyze the performance of DPSR in terms of quality of the super-resolved images as well as computational complexity with respect to several state-of-the-art hyperspectral SR models on multiple datasets. Code and pretrained models will be available at: \url{https://github.com/DavidePiccinini98/dpsr}.

\subsection{Implementation Details and Experimental Settings}
The proposed architecture makes use of an internal feature dimension $F$. This hyperparameter was set to $F=280$ for all experiments except those where we tested scaling down the model to $F=128$ (DPSR-S model). The expansion factor of the Mamba Block was set to $E=1$, whereas the internal state of the SSM used was $N=16$, following \cite{gu2023mamba} and common practice in the literature that builds on the Mamba model. The reduced features factor in the Upsampler which controls how much we compress the feature information while recovering the right spatial resolution was set to $f=64$ throughout all the experiments. For the compact DPSR-S variant, our target was real-time operation on low-power hardware; this requirement drove the choice of internal parameters. For the full DPSR model, both the internal feature width and the Mamba expansion factor were selected to match a predefined complexity envelope (30K FLOPs/px on HySpecNet-11k with a SR factor of 4 and 20K FLOPs/px on the same dataset with a SR factor of 2).
For the training procedure, we used the Adam optimizer with a fixed learning rate of $10^{-4}$, and a variable number of epochs to let the model converge until overfitting was observed: the optimal number of training epochs depends on the dataset and it amounts to 300 for Pavia, 1600 for Houston, and 2000 for HySpecNet-11k and Chikusei. This value was determined empirically by monitoring convergence and stopping training once the improvement in validation MPSNR fell below 0.01 dB over the preceding 50 epochs. No weight decay was used. We used the same loss as in \cite{chen2024cross}, setting in the same way $\alpha_{s}$ to $0.3$ and $\alpha_{g}$ to $0.1$, as common practice for other methods in the literature. 

We present results on four different hyperspectral datasets. Besides the widely used Chikusei \cite{yokoya2016airborne}, Houston \cite{le20182018} and Pavia \cite{benediktsson2005classification}, we add the HySpecNet-11k dataset \cite{fuchs2023hyspecnet}. This is motivated by the significantly larger number of images present in HySpecNet-11k which makes the results less sensitive to overfitting. See Section \ref{sec:hsi_results} for details regarding the preprocessing of each of the four datasets.
Since our model deploys an interpolation strategy, i.e., super-resolves the previous line, in order to train and test with SR factor $r$, we discard the first $r$ lines of the output of the network (as they estimate lines before the start of the image) and the last $r$ lines of the ground truth (as these cannot be estimated by DPSR as it would require a line after the end of the image). Notice that in a real-world scenario, we would not have to discard any line since the stream of input lines would be continuous. 
For the sake of fairness, we use the same discarding approach for all the other baselines we compare against in the evaluation of quality metrics.

We compare DPSR against the following state-of-the-art and baseline methods: CST \cite{chen2024cross}, MSDformer \cite{chen2023msdformer}, SNLSR \cite{hu2024exploring}, EUNet \cite{liu2023efficient}, SSPSR \cite{jiang2020learning}, GDRRN \cite{li2018single}. The official code released by the authors has been used to retrain and test the models, with all hyperparameters set as specified in the published reference papers. We remark that we do not report results for SNLSR at $2\times$ SR factor since the model has been intrinsically designed and proposed for the $4\times$ factor. In particular, the architecture uses a two-stage spatial expansion, each building its upsampling module with a scale equal to 2. For the method to support a true $2\times$ SR factor major modifications to the model architecture and size would be required, which would be unfair to the method as proposed by the authors. Additionally, we report baseline results obtained using bicubic interpolation when processing the entire image.

To quantify the fidelity of the super-resolved hyperspectral images, we report four commonly used metrics: peak signal–to–noise ratio (PSNR) \cite{mannos1974effects}, structural similarity (SSIM) \cite{wang2004image}, spectral‐angle mapper (SAM) \cite{yuhas1992discrimination}, and root‐mean‐squared error (RMSE). Together, these four metrics provide a balanced view of spatial quality (PSNR, SSIM), spectral faithfulness (SAM), and overall error magnitude (RMSE). Following \cite{chen2023msdformer, chen2024cross}, for a hyperspectral image with $C$ spectral channels, we compute each metric channel-wise, accumulate over all the channels and then take the average as the metric value.

For model training and testing we used one NVIDIA TITAN RTX GPU, whereas for the low-power inference speed experiments, we used an NVIDIA Jetson Orin Nano in $15$ W mode.

We remark that all reported results were obtained with a fixed random seed for every model and dataset, for both $2\times$ and $4\times$ settings, and for all ablation studies, thus representing a typical run, as commonly done in the literature.

\subsection{Experimental Results on Hyperspectral Datasets}\label{sec:hsi_results}
We first conducted experiments on the HySpecNet-11k dataset  \cite{fuchs2023hyspecnet}, a large-scale hyperspectral benchmark dataset consisting of 11,483 non-overlapping image patches. Each patch has $128 \times 128$ pixels, with 202 usable spectral bands, and ground sampling distance of 30 m. The standard training set comprises 70\% of the patches, the validation set includes 20\% of the patches, and the test set contains the remaining 10\%.
Moreover, the dataset provides two types of splits: the easy split, where patches from the same tile can appear in different sets (patchwise splitting), and the hard split, where all patches from a single tile are restricted to the same set (tilewise splitting). 

All experiments in this work were conducted using the hard split configuration. The use of this new dataset with a sensor with a larger number of spectral channels as well as a significantly larger number of images compared to traditionally-used dataset, such as Pavia, allows for more reliable model testing, avoiding the risk of overfitting which afflicts the vast majority of architectures on the smaller datasets. 

We remark that approximately $8\%$ of the images that are provided in the HySpecNet-11k dataset present some zeroed-out channels, with variability in which channels present this behavior. In our experiments, these channels were not excluded during training and used as they are, but they were discarded for computation of quality metrics during testing, as they represent unreliable and unrealistic predictions.   
We also used classic augmentations in order to increase the number of training samples by a factor of 8. 
We employed bicubic downsampling on the images before feeding them into the models, resulting in inputs with dimensions $32 \times 32 \times 202$ for SR factor $r=4$ and $64 \times 64 \times 202$ for $r=2$. 

We also conducted experiments on Chikusei \cite{yokoya2016airborne}, Houston \cite{le20182018} and Pavia \cite{benediktsson2005classification} datasets. Regarding Chikusei and Houston datasets, we followed exactly the same pipeline employed in \cite{chen2023msdformer} for cropping each original image and obtain the train, validation and test splits for each of the two. With respect to the Pavia dataset, we followed the cropping and splitting strategy introduced in \cite{chen2024cross}. It is important to note that this last choice was due to the fact that CST requires specific spatial sizes to work: in \cite{chen2023msdformer} the test images utilized have spatial sizes $224 \times 224$, incompatible with the CST architecture, whereas in \cite{chen2024cross} the authors used $256 \times 256$ test crops. For all four datasets, no band selection or denoising was applied; however, images were normalized to the $\left[ 0,1 \right]$ range for both training and testing. 

Table \ref{tab:hysp_x4} presents the results on HySpecNet-11k in the $4\times$ SR setting and also summarizes the complexity of all methods in terms of number of trainable parameters and number of floating point operations per image pixel. We first remark how the $31K$ FLOPs/pixel of DPSR are significantly smaller than that of state-of-the-art methods, often by an order of magnitude. 

Despite this, we notice how DPSR delivers image quality close to that of significantly more methods. Indeed, it is only 0.28 dB and 0.44 dB shy of the complex Transformer-based MSDformer and CST models, respectively, while outperforming all the other recent state-of-the-art models. Similar considerations hold for the $2\times$ SR factor on HySpecNet-11k presented in Table \ref{tab:hysp_x2}.
Regarding the results obtained on Chikusei, Table \ref{tab:chiku_x4} and Table \ref{tab:chiku_x2} show that DPSR essentially matches MSDformer and comes within 0.1 dB of CST PSNR. 
For the Houston urban scene imagery, Table \ref{tab:houston_x4} and Table \ref{tab:houston_x2} show that the reconstruction fidelity of DPSR is again very close to that of the leading models on both SR factors.
Finally, the Pavia benchmarks presented in Table \ref{tab:pavia_x4} and Table \ref{tab:pavia_x2} also confirm the trend observed on the other datasets, highlighting the excellent complexity-quality tradeoff offered by DPSR.

Figures \ref{fig:test_247} and \ref{fig:test_628} show a qualitative comparison of the methods on two test scenes from the HySpecNet-11k dataset and confirms that no visual artifacts are introduced by the line-based approach. This is also confirmed by inspecting Figures \ref{fig:test_247_error} and \ref{fig:test_628_error}, which show the error spectra of every method on each of the two HySpecNet-11k test images. Figure \ref{fig:absolute_error} reports the per-band mean absolute error on the test set of the Houston dataset. It is clear that DPSR is almost as accurate as CST in every band.

\subsection{Analysis of Memory and Computational Efficiency}

We now focus on better investigating the efficiency of the proposed DPSR model with respect to state-of-the-art baselines. In particular, we first analyze the memory requirements of each method and how they scale with image size. Figures \ref{fig:memory_dims_x4} and \ref{fig:memory_dims_x2} report GPU memory usage as a function of each image dimension, while the other two are fixed to values that are representative of real images acquired by satellites. The results clearly illustrate the efficiency of our model. First, we notice how DPSR consistently requires 4 to 8 times less memory than other methods. In absolute terms, an image of size $1000 \times 1000 \times 66$ which constitutes a frame of the PRISMA VNIR instrument can be processed by DPSR with less than 1GB of memory, a size compatible with current low-power computing platforms. On the contrary, state-of-the-art models require more than the 24GB of VRAM available on the GPU used for this experiment, so much so that this point is not reported in the previous figures. We also remark that memory usage is constant with the number of lines by design. While it grows linearly with the number of columns, this dependency is very weak and we do not see this trend in the figures due to the value being very small and dominated by other implementation overhead.

\begin{figure}[t]
    \centering
    \includegraphics[width=\linewidth]{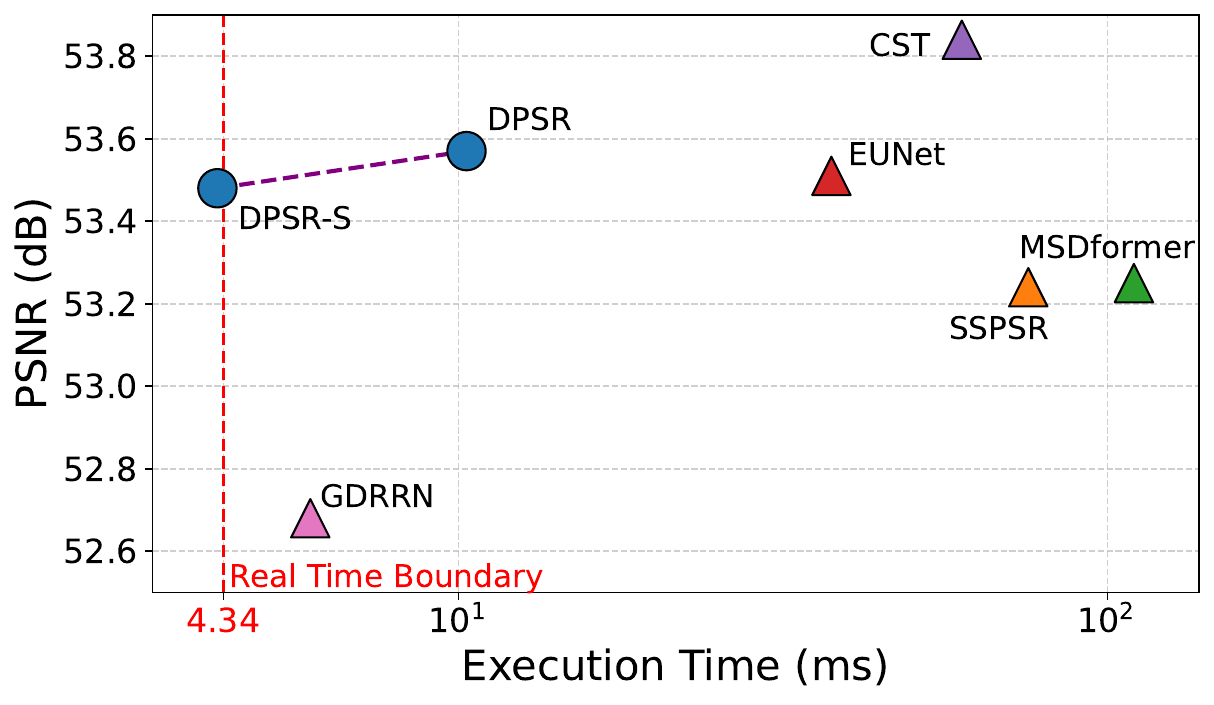}
    \caption{Line processing time ($W=1000$, $C=66$) on Nvidia Jetson Orin Nano and PSNR on Houston for $2\times$ SR.}
    \label{fig:speed_x2}
\end{figure}
\begin{figure}[t]
    \centering
    \includegraphics[width=\linewidth]{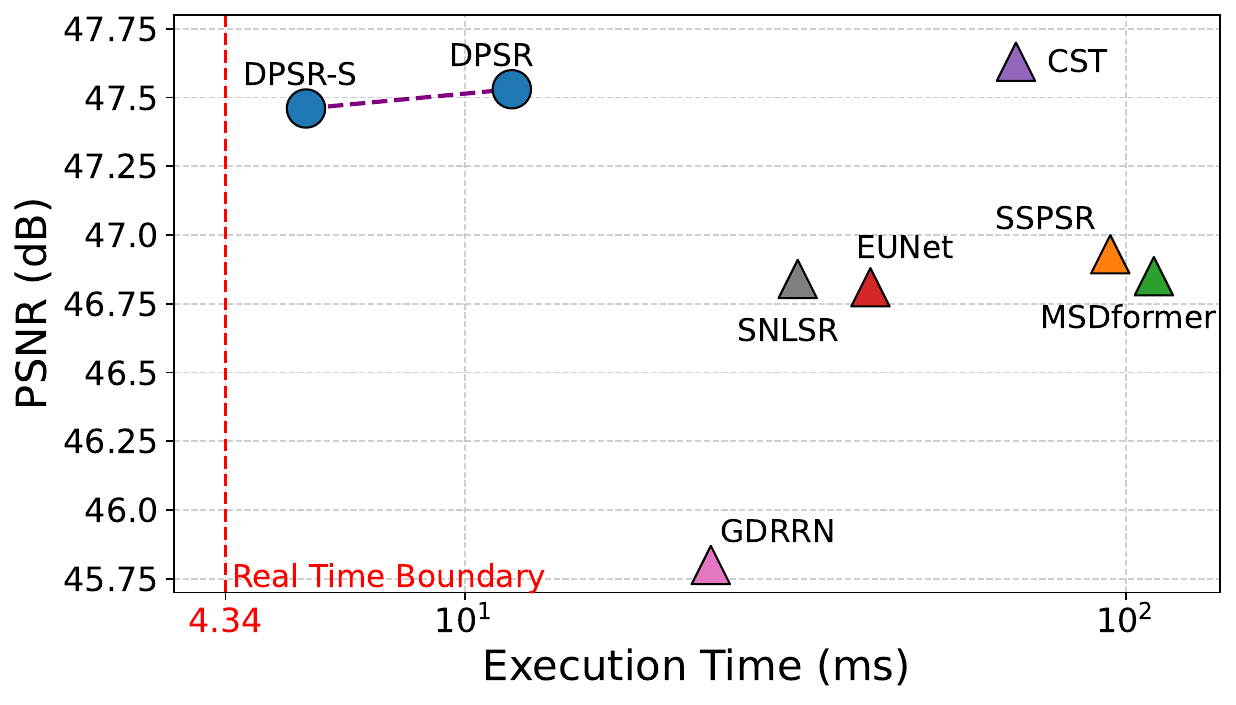}
    \caption{Line processing time ($W=1000$, $C=66$) on Nvidia Jetson Orin Nano and PSNR on Houston for $4\times$ SR.}
    \label{fig:speed_x4}
\end{figure}

In addition to testing memory requirements, we also characterize the performance of the method in terms of inference speed on a low-power platform, namely the NVIDIA Jetson Orin Nano which requires only $15$ W of power for the full system.
We benchmark every model in half-precision floating point, but without further optimizations. We use a test input of size with 1000 columns and 66 bands. We present results on two variants of DPSR, namely the one with $F=280$ features reported in the previous experiments (DPSR) and a smaller version with $F=128$ (DPSR-S). 
Figures \ref{fig:speed_x2} and \ref{fig:speed_x4} present the resulting runtime normalized by the number of lines processed to have an equivalent line processing time. This allows a comparison with the actual system time needed to acquire a line by existing pushbroom sensors to determine whether a method can perform super-resolution in real time. For example, the PRISMA VNIR instrument acquires a  $1 \times 1000 \times 66$ every 4.34ms \cite{prismaATBD}. We can see that DPSR-S requires 4.25ms for $2\times$ SR and 5.70ms for $4\times$ SR, thus satisfying or being marginally above the real-time requirements. DPSR requires a slightly longer 10.28ms and 11.76ms, respectively, but it is still substantially faster than state-of-the-art methods. We remark that our current implementation is not yet hardware-optimized. While the unoptimized DPSR-S already satisfies real-time speed requirements, the base model falls slightly short, though it still outperforms state-of-the-art methods by a significant margin. Future work will explore hardware-aware optimizations, such as kernel-level tuning and FPGA deployment, to further enhance throughput. At the same time, we notice that state-of-the-art methods exhibit significantly higher runtimes which do not allow to get close to real-time processing.

\subsection{Ablation Experiments}
\label{subsec:ablation}
In this section, we validate some design choices of our model through ablation experiments. Throughout the tables, Parameters, FLOPs/px, and Memory are computed on inputs of size 1 × 1000 × 66 to emulate onboard sizes of the PRISMA satellite, in order to provide consistency across different eperiments and to simulate onboard deployment metrics.

We first assess the effectiveness of the memory mechanism implemented by Mamba by comparing it against alternative approaches. In particular, we consider alternatives based on simple causal convolution, an LSTM recurrent neural network \cite{hochreiter1997long}, and a recently proposed improvement over Mamba (Mamba-2) \cite{dao2024transformers} with expansion factor of $1$, head dimension of $10$, number of heads of $28$ and number of groups of $1$, using its multi-input SSM version. 
 
\begin{table}[t]
\centering
\caption{Architecture ablation.  MPSNR is computed on the test set of the HySpecNet-11k dataset with SR factor of $4$. Parameters and FLOPs/px are computed on inputs of size 1 x 1000 x 66, reflecting the acquisition dimensions of PRISMA VNIR instrument, in order to simulate onboard deployment.}
\label{tab:causalconv}
\begin{tabular}{cccc}
\textbf{Model} & \textbf{MPSNR (dB)} & \textbf{Params} & \textbf{FLOPs/pixel} \\
\hline\hline
CausalConv & 42.96 & 2.68 M & 138.78 K \\
LSTM       & 43.05 & 3.31 M & 100.90 K \\
Mamba     & \textbf{43.17} & 2.57 M & 78.18 K \\
Mamba-2    & 43.12 & \textbf{2.55 M} & \textbf{77.72 K} \\
\hline
\end{tabular}
\end{table}

\begin{table}[t]
\centering
\caption{Model size ablation. The numbers reported represent MPSNR (dB) under two super-resolution factors across four datasets.}
\label{tab:model_size2}
\begin{tabular}{cccccc}
\textbf{SR Factor} & \textbf{Features} & \textbf{HySpecNet-11k} & \textbf{Chikusei} & \textbf{Houston} & \textbf{Pavia} \\
\hline\hline
$2\times$ & $F=128$ & 48.35 & 47.36 & 53.48 & 35.80 \\
$2\times$ & $F=280$ & \textbf{48.85} & \textbf{47.49} & \textbf{53.57} & \textbf{35.87} \\
\hline
$4\times$ & $F=128$ & 42.80 & 40.07 & 47.46 & 29.00 \\
$4\times$ & $F=280$ & \textbf{43.17} & \textbf{40.07} & \textbf{47.53} & \textbf{29.05} \\
\hline
\end{tabular}
\end{table}

\begin{table}[t]
\centering
\caption{Comparison with LineSR \cite{piccinini2025LineSR}.MPSNR is computed on the test set of the HySpecNet-11k dataset with SR factor of $4$. Parameters, FLOPs/px and Memory are computed on inputs of size 1 x 1000 x 66, reflecting the acquisition dimensions of PRISMA VNIR instrument, in order to simulate onboard deployment. }
\label{tab:linesr}
\begin{tabular}{ccccc}
\textbf{Model} & \textbf{MPSNR (dB)} & \textbf{Params} & \textbf{FLOPs/pixel} & \textbf{Memory} \\
\hline\hline
LineSR     & 41.82 & \textbf{95.11 K} & 173.78 K & 2684.62 MiB \\
\textbf{DPSR} & \textbf{43.17} & 2.57 M & \textbf{78.18 K} & \textbf{204.58 MiB} \\
\hline
\end{tabular}
\end{table}

These approaches replace the Mamba Block in our architecture while retaining the rest of the architecture, as a way to measure just the effectiveness of the line memory mechanism. Table \ref{tab:causalconv} shows the results of this experiment, which confirms that Mamba outperforms other approaches. Mamba-2 does not seem to provide significant improvements over the original Mamba for our setting, so we retain the simpler Mamba design.

Moreover, Table \ref{tab:model_size2} 
reports how the the proposed architecture scales with respect to the number of features $F$ across all datasets. We compare the DPSR-S ($F=128$) model whose unoptimized implementation already reaches real-time performance and the base DPSR ($F=280$) model, which we conjecture could reach real-time speed with some optimized implementation. Overall, we notice that the smaller model is still quite close to state-of-the-art baselines in terms of image quality.

Table \ref{tab:linesr} shows a comparison between DPSR and a preliminary design (LineSR) which we proposed in \cite{piccinini2025LineSR}. The results clearly show that DPSR yields significantly better results at reduced computational and memory requirements.

Then, we ablate the effect of the internal expansion factor of the Mamba Block in Table \ref{tab:exp_factor}. We chose to fix  $E=1$ in order to optimize the memory and power requirements and also in order not to damage the inference speed, given the tradeoff between performance gain and efficiency drop: in particular, an expansion factor $E=2$ would require $16.5\% $ more FLOPs/px and $18.1\%$ more memory while delivering a negligible quality improvement in terms of MPSNR on the test set of the Houston dataset: $47.55$ dB instead of $47.53$ dB.     

\begin{table}[t]
\centering
\caption{Impact of Mamba internal expansion factor E on the results of DPSR. MPSNR is computed on the test set of the Houston dataset with SR factor of $4$. Parameters, FLOPs/px and Memory are computed on inputs of size 1 x 1000 x 66, reflecting the acquisition dimensions of PRISMA VNIR instrument, in order to simulate onboard deployment. Remember that throughout the paper DPSR makes use of $E=1$.}
\label{tab:exp_factor}
\begin{tabular}{ccccc}
\textbf{Model} & \textbf{MPSNR (dB)} & \textbf{Params} & \textbf{FLOPs/pixel} & \textbf{Memory} \\
\hline\hline
$E=1$ & 47.53 & \textbf{2.57 M} & \textbf{78.18 K} & \textbf{204.58 MiB} \\
$E=2$ & \textbf{47.55} & 3.09 M & 93.63 K & 249.91 MiB \\
\hline
\end{tabular}
\end{table}

\begin{table}[t]
\centering
\caption{Impact of bilinear residual on the results. MPSNR is computed on the test set of the Houston dataset with SR factor of $4$. Parameters, FLOPs/px and Memory are computed on inputs of size 1 x 1000 x 66, reflecting the acquisition dimensions of PRISMA VNIR instrument, in order to simulate onboard deployment.}
\label{tab:bilinear}
\begin{tabular}{ccccc}
\textbf{Model} & \textbf{MPSNR (dB)} & \textbf{Params} & \textbf{FLOPs/pixel} & \textbf{Memory} \\
\hline\hline
No-Res DPSR & 47.44 & 2.57 M & 78.18 K & \textbf{204.50 MiB} \\
\textbf{DPSR} & \textbf{47.53} & 2.57 M & 78.18 K & 204.58 MiB \\
\hline
\end{tabular}
\end{table}

\begin{table}[t]
\centering
\caption{Comparison between the usage of bilinear and bicubic residual. MPSNR is computed on the test set of the Pavia dataset with SR factor of $4$. Parameters, FLOPs/px and Memory are computed on inputs of size 1 x 1000 x 66, reflecting the acquisition dimensions of PRISMA VNIR instrument, in order to simulate onboard deployment. Remember that throughout the paper DPSR makes use of the bilinear upsampling.}
\label{tab:bicubic}
\begin{tabular}{ccccc}
\textbf{Model} & \textbf{MPSNR (dB)} & \textbf{Params} & \textbf{FLOPs/pixel} & \textbf{Memory} \\
\hline\hline
Bilinear & 29.05 & 2.57 M & 78.18 K & \textbf{204.58 MiB} \\
Bicubic  & 29.05 & 2.57 M & 78.18 K & 205.91 MiB \\
\hline
\end{tabular}
\end{table}

We show in Table \ref{tab:bilinear} that allowing the network to learn a residual correction of the bilinear upsampling delivers better performance, i.e. around $0.1 dB$, with a negligible parameter and memory overhead.

Finally, in Table \ref{tab:bicubic} we evaluate the impact of bilinear and bicubic upsampling. Notice that the bilinear interpolator operates using only the current LR line $y$ and the previous line $y-1$, whereas bicubic interpolation requires buffering at least three earlier lines. We notice that in this experiment bilinear and bicubic residuals achieve the same MPSNR. Since there are no substantial quality benefits, we opt for the simpler and lower-memory bilinear upsampling.

\section{Conclusions}
Overall, DPSR proves that respecting the acquisition physics, i.e., pushbroom, causal access, could be a successful strategy to design neural networks suited for onboard deployment, making such deep learning models strong candidates to be the future of onboard processing systems. Moreover, exploiting selective SSM memory can recover much of the accuracy of heavier 2D models at a fraction of their cost. While a small fidelity gap remains relative to the best-performing state-of-the-art, the ability to operate within a single line time and with minimal memory budgets strongly supports the viability of onboard, real-time HSI super-resolution.

\small
\bibliographystyle{IEEEtran}
\bibliography{references}

\end{document}